\documentclass[12pt]{article}
\usepackage{latexsym,amsmath,amssymb,amsbsy,graphicx}
\usepackage[cp1251]{inputenc}
\usepackage[T2A]{fontenc}
\usepackage[russian,english]{babel}
\usepackage[space]{cite} 

\begin{document}
\bigskip

{\bf {New classification parameter of solar flares based on the
maximum flux in soft X-rays and on duration of flare}}

\bigskip

\centerline {E.A. Bruevich}

\centerline {\it Lomonosov Moscow State
 University, Sternberg Astronomical Institute,}
\centerline {\it Universitetsky pr., 13, Moscow 119992, Russia}\

\centerline {\it e-mail:  {red-field@yandex.ru}
 }\

\bigskip

{Abstract.} Solar flare activity is characterised by different classification
systems, both in optical and X-ray ranges. The most generally
accepted classifications of solar flares describe important
parameters of flares such as the  maximum of brightness of the flare
in the in optical range - $H_{\alpha}$ flare class 
(change from F to B), area of the flare in  $H_{\alpha}$ 
(change from S for areas less than 2 square degrees to 4 for areas more than 24.7
square degrees) and the maximum amplitude of the soft X-ray (SXR-flux) in
the band 0.1 -- 0.8 nm ($F_{0.1-0.8}^{max}$) -- X-ray flares of classes from C to X. 
A new classification parameter of solar flares is proposed here -- the X-ray 
index of flare XI, based on GOES measurements of solar radiation in the SXR-range. The XI-index has a clear
physical interpretation associated with the total flare energy in
the SXR-range. XI is easily calculated for each flare with
use of available GOES data. The XI-index can be used  along with other geoeffective parameters of Solar activity to assess both flares and  Coronal mass ejections (CMEs ) that are connected with them.

\bigskip

\vskip12pt
\section{Introduction}
\vskip12pt

Solar flares  and coronal mass ejections (CMEs) are the most powerful manifestations of solar activity.
Solar flares are a complex of physical phenomena in plasma combined
into one interconnected process of energy accumulation and release.
Flares manifest themselves in all ranges of the electromagnetic
spectrum, that makes it possible to study the physical processes
occurring in them.

Solar perturbations of the explosive type have become more accessible for analysis
in recent years with the development of space technology, and in particular through the regular X-ray observations on the GOES series satellites and  in the ultraviolet range on the SDO orbital observatory. Solar activity of the  explosive type are accompanied by CMEs with powerful release of energy, primarily in the form of kinetic movements of plasma (shock waves, coronal emissions of matter), as well as in the form of enhanced in flares fluxes of electromagnetic radiation, solar wind and accelerated particles. Flare fluxes in  Ultraviolet and X-ray ranges dramatically increase the ionization in the upper atmosphere of the Earth and in the ionosphere. All these  processes take place against the background of the perturbed interplanetary magnetic field.  Particles of high energies from the flares and from the CMEs assosiated with them penetrate the upper atmosphere of the Earth with destroying the ozone layer. Shock waves and solar plasma emissions after large
flares cause severe disturbances of the magnetosphere of the Earth -- the magnetospheric storms.
 Each of these factors has different effects on the near-earth space, that can lead to violations of radio communication, malfunctions in
the navigation devices of ships and aircrafts and radar systems. The largest flares are sometimes accompanied by the
so-called ground level events -- GLE (Ground Level Enhancement). GLE -- a rather rare and outstanding event.

There is a rather low flare activity in the current cycle 24 
(Bazilevskaya {\it{et al.}} 2015; Bruevich \& Yakunina 2017). 
A comparative analysis of several solar activity
indices in cycles 22, 23 and 24 showed that the relative differences
in the amplitudes of variations of activity indices from the minimum
to the maximum of the cycle vary significantly during the transition
from cycles 22 and 23 to cycle 24. But, the maximum amplitudes of
variations of the flare index( FI), the relative number of sunspots
SSN and in the UV flux in the hydrogen line 121.6 nm $F_{L \alpha}$
are significantly decreased (by 20 -- 50\%) already in cycle 23.

When analysing solar flares, all the parameters characterising this phenomenon are important: the area of the flare, its average
brightness, the shape of the light curves as in the optical range
so and in the ultraviolet and X-ray ranges. Both the study of maximum amplitudes
of the flare in different bands and lines of the emission spectrum
and also the study of the total energy that came from the flare to the Earth are very
important (Bruevich \& Bruevich 2018).

Currently, two classification systems are used to determine the
flare class: (1) -- the optical classification (flares class in
$H_{\alpha} $ changed from SF to 4B), supplemented by the flare
index FI (considering the full duration of the flare in minutes) and
(2)-- an X-ray classification based on the absolute maximum of the
flare flux in SXR-range -- (changed from
C1 to X27). In the X-ray classification, the flare duration and the
shape of the X-ray luminosity curve are not considered.
Further, the radiation fluxes are measured in $W / m^2$.

The energy accumulation  in the form of magnetic energy of the current layer in the upper chromosphere and in the corona occurs in the active region before the flare. The current layer has a
magneto-plasma structure, at least two-dimensional structure and
usually two-scale structure. The first, who attracted attention to the
importance of the process of formation of the current layer in the
corona near a special line of the magnetic field, was S. I.
Syrovatsky (Somov \& Syrovatskii 1976).

A huge energy is suddenly released to the beginning of a large flare
at the top of the arch of the magnetic field according to the
present ideas about the development of the flare process: in the
SXR-region the flare radiation is orders of magnitude higher
than the radiation of the solar disc without of flares in this
range. Around the primary energy release area, electrons (and sometimes
protons) are accelerated to high energies and plasma is heated to temperatures from 20 to 30 million K.

The problem of predicting the occurrence of large solar flares is
actively studied in spectral-polarization observations of the Sun in
the microwave range. On the RATAN-600 radio telescope 
the continuous monitoring of the flare-productive active regions on
the Sun that can generate powerful X-ray flare events of M and X
classes is carried out.

These observations show the existence of a rather long preceding
phase in the pre-flare emission of the active regions. This phase,
that precedes to the flares of large power, is characterised by the
rise of a new magnetic flux and by multiple inversions of the sign
of circular polarization in the wavelength range from 2 to 5 cm (Bogod 2006).

Such unique opportunities in flares forecasting relate to the 
improvement of parameters of the multi-octave solar
spectral-polarization complex of high resolution (SPCHR) on
RATAN-600 and with the successful implementation of the Program of
Regular Observations and Data Processing using modern technologies (Bogod 2011).

Thus, the beginning and further development of the flare in
different spectral ranges (and different spectral lines of the
short-wave part of the spectrum formed both in the chromosphere and
in the corona) occurs in different ways, so for a more complete
description of the observed flare parameters, both optical and X-ray
classifications are used.

An important point for determining the flare class in the optical
classification is the identification of letters and numbers with the
real parameters of the X-ray classification based on the magnitude
of the fluxes at the flare maximum (Ozgus {\it{et al.}} 2003).

In Kleczek (1952), the value determining the flare index in the optical range
$Q = it$ is proportional to the total energy emitted by the flare
was introduced for the first time. In this equation i represents the
class of $H_{\alpha}$-flares in a special scale, t defines the
duration of the $H_{\alpha}$-flare in minutes. The value of i
changes from 0.5 for SF, SN and SB flare, to 4.0 for the 4B
$H_{\alpha} $-flares. Now for Q received the  FI (Flare Index)
designation, (Ozgus {\it{et al.}} 2003; Kleczek 1952).

The FI value is calculated as the averages of the day and adjusted
for the total observation time during the day. The archived data of
FI from 1976 to 2014 are available on the web-site of the National
Geophysical Data Center -- NOAA, avalable at www.ngdc.noaa.gov/stp/space-weather/solar-data/solar-features/solar-flares/index/flare- index/.

The system of evaluation of solar flares power by X-ray radiation
(classes C, M and X) adopted all over the world currently relies on
measurements of radiation flux in SXR-region, see,
for example (Altyntsev {\it{et al.}} 1982; Bowen {\it{et al.}} 2013a; Bowen {\it{et al.}} 2013b). The most powerful flares in this
classification -- flares of X class corresponds to the absolute flux
of more than $10^{-4} W / m^2$ in the SXR-range, X-ray
flares of M1 -- M9 classes corresponds to the flux from $10^{-5}
W/m^2$ to $10^{-4} W/m^2$, X-ray flares C1 -- C9 corresponds to the
flux from the $10^{-6} W/m^2$ to $10^{-5} W / m^2$.

\begin{figure}[tbh!]
\centerline{
\includegraphics[width=115mm]{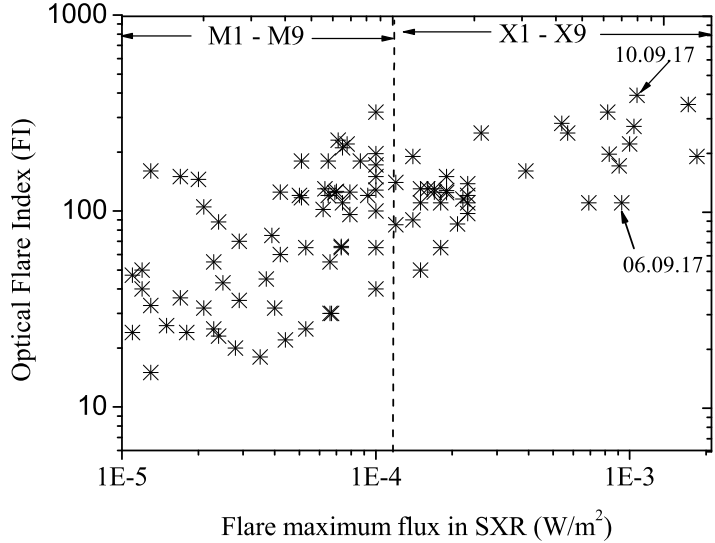}}
 \caption{The relationship between optical flare index (FI) and X-class of flares (peak flux from the flare
 in the range 0.1 -- 0.8 nm -- $ F_{0.1-0.8}$) for 96 flares of cycles 23 and 24. The vertical dotted line separates classes
 M and classes X of the flares.}
{\label{Fi:Fig1}}
\end{figure}

\section{Comparison of flares in the parameters according to the optical and X-ray classifications}
{\label{S:place}}

Figure 1 demonstrates (for 96 flares of cycles 23 and 24 from the
Table 1 lower) the relationships between the two most common flare's
classification systems -- the optical classification (expressed in
the FI index value) and X-ray classification, based on the GOES data
in the SXR-range.

 It can be seen that the amplitude of the flare flux $F_{0.1-0.8}^{max}$ and the FI optical index, that is the energy analogue
 of the flare in $H_\alpha$, do not show a noticeable relationship although on average the greater value of $F_{0.1-0.8}^{max}$ 
 corresponds to the larger value of FI. It can be noted that for flares of X-ray class X1 -- X9, the FI optical index varies from 50 to 700, while for M1 -- M9 classes FI varies from 10 to 300.

It can be seen from Figure 2 that the relationship between the FI optical flare index and the total energy $E_{0.1-0.8}$ calculated in this work for the flares from Table 1 according to formula (1) is closer than the relationship in Figure 1.

Both flare indices that are shown in Figures 1, 2 are a measure
of the energy emitted in the optical and X-ray bands. Note that the
duration of the flare in the SXR-range is determined quite
accurately, as the data of the GOES measurements are presented with
a time interval of 2.5 seconds.

In the optical range, the duration of the flare and its optical
score can vary between observations at different observatories (up
to tens of minutes for long-duration flares) that creates an
additional error in the definition of FI.

To determine the FI values related to individual flares, we used
information from the Catalogs, published in (The Catalogue 2008; Preliminary Current Catalogue 2018 ), where unfortunately, 
the moments of the end of flares were not always
determined with sufficient accuracy.

As for the X-ray classification, the example for flares 9.03.11 and
12.07.12 (Figure 3 and Figure  4) illustrates its wrongness although
this X-ray classification is used most frequently. These flares are
about the same X-ray class (equal to X1.6 and X1.4) but their
total energy emitted in this range varies greatly: $E_ {0.1-0.8} =
0.15 ~J/m^2$ for the flare 9.03.11 of X1.6 class and $E_{0.1-0.8} =
0.75 ~J/m^2$ for the flare 12.07.12 of X1.4 class, respectively.

Note that the total energy $E_{0.1-0.8}$ emitted by the flare is
calculated in this work using the time integration of the flux
$F_{0.1-0.8} (t)$ given the values of the background radiation
$F_{background}$ from the beginning to the end of the flare:

$$ E_{0.1-0.8} =\int (F_{0.1-0.8} (t) - F_{background}) dt   ~~~~~~~~(1)  $$

Such a significant difference in the total energy of 9.03.11 and
12.07.12 flares is explained by the difference in the shape of the
curves of brightness and duration of flares. Along with the X-ray
class, its optical class (from SF to 4B) is used to describe the
flare. To determine the optical class of the flare, you need to have
access to observations of both the flare's area and its brightness
in the $H_{\alpha}$ line.

The FI flare activity index complements the information about the
area and brightness of the flare with information about its duration
in the optical range. The difficulty in calculating FI is that the
flare duration in the optical range may differ for observations in
different observatories. In this sense, the GOES series of
observations available in real time and with common absolute
calibration (that allows comparison of flare events since 1978) have
a huge advantage over the classification in optical range: no wonder
that the X-ray classification based only on knowledge of the
amplitude in the flare maximum is currently the most popular.

The most powerful flares of 24th cycle that occurred at the decline
phase of cycle during September 2017 confirm that the
classification based only on the maximum amplitude value does not
carry complete information.

According to this classification, the flare on September 6, 2017 of
class X9.3 is considered to be more powerful than the X8.2 flare of
10 September 2017, originating from the same active region. In fact,
the flare X8.2 was much stronger than the flare X9.3, the curve of
dependence of flare power versus time was kind of more gently
sloping and so the total energy of X8.2 flare $E_ {0.1-0.8} $  was
much greater than for X9.3 flare ($2.52 ~J/m^2$ vs $0.35 ~J/m^2$
respectively).

Related to this is the fact important for the effect of flares on
the magnetosphere and the ionosphere: a stream of protons ($I_{pr}$)
in the channel with E $\geqslant $ 10 MeV caused by flare X8.2, was much
higher than after flare X9.3. For the more hard-energy protons with
E $\geqslant$ 100 MeV the flare X 9.3 there was virtually no increase in the
flux above the background level. But in flare X8.2 the strengthening of
the proton flux reached a record for the 24th cycle value.

\begin{figure}[tbh!]
\centerline{
\includegraphics[width=115mm]{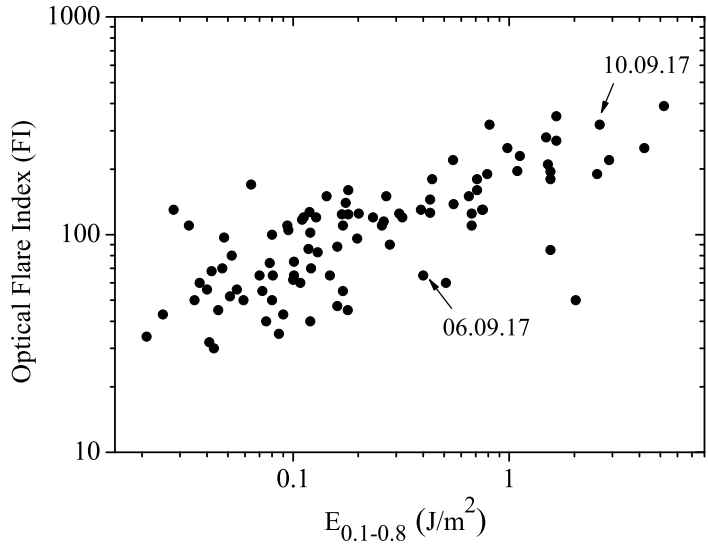}}
 \caption{ The connection between the optical flare index (FI) and the total flare energy of
 $E_ {0.1-0.8}$ in the SXR-range for 96 flares of 23rd and 24th cycles.}
{\label{Fi:Fig2}}
\end{figure}

\section{The XI -- new X-ray solar flares index, determined from observations on the satellites of the GOES series}
{\label{S:place}}

The most clear option for a physically justified classification of
flares is the addition of the information about flare's duration to
the X-ray classification of flares in terms of the maximum SXR-flux.

Thus, by analogy with the value of the optical flare index FI
(proportional to the total energy radiated in $H_{\alpha}$), X-ray
flare activity parameter XI (XI-index) based on GOES data (considering the duration
of the flare in the X-ray range and the shape of the flare light
curve) is introduced in this paper. The new flare parameter XI is also
an analogue of the total energy $E_{0.1-0.8}$.

\begin{figure}[tbh!]
\centerline{
\includegraphics[width=115mm]{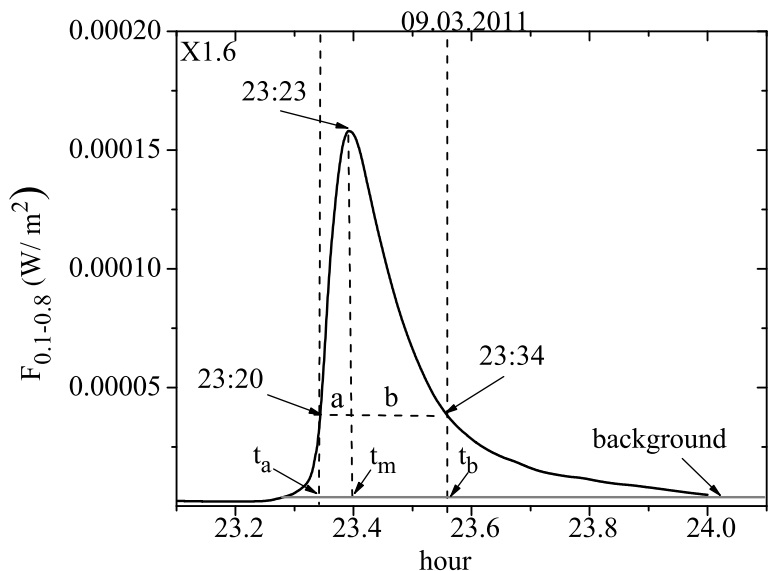}}
 \caption{Flare 09.03.2011. The radiation flux in the SXR-range $F_{0.1-0.8}$ according to the observations of GOES-15.
 The moments of the maximum of flare flux  $t_m$ and moments of a quarter of the maximum flux value (FWQM) -- $t_a$ and $t_b$ are showed.}
{\label{Fi:Fig3}}
\end{figure}

\begin{figure}[tbh!]
\centerline{
\includegraphics[width=115mm]{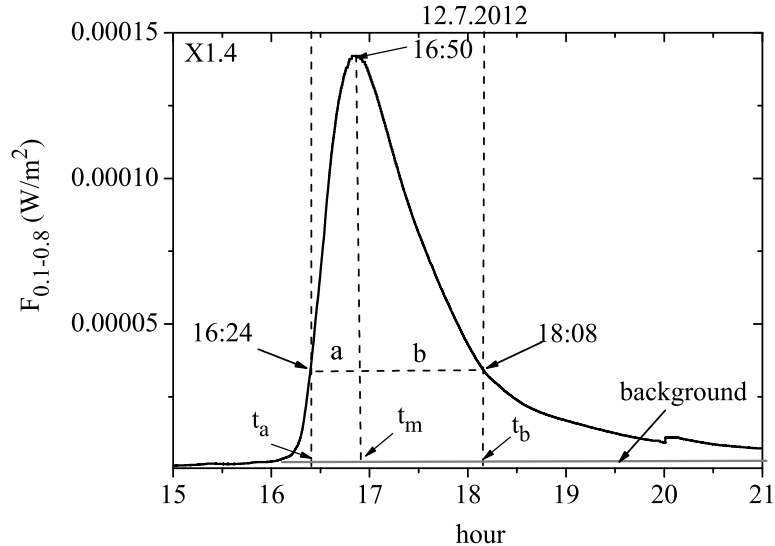}}
 \caption{Flare 12.07.2012. The radiation flux in the SXR-range  $F_ {0.1-0.8} $ according to the observations of GOES-15.
 The moments of the maximum of flare flux $t_m$ and moments of a quarter of the maximum flux value (FWQM) -- $t_a$ and $t_b$ are showed.}
{\label{Fi:Fig4}}
\end{figure}

To determine XI, we use the value of  quarter of the maximum flux
(FWQM - full width quarter-maximum). In Figure 3 and
Figure 4, the value of a corresponds to the time in minutes that has
elapsed from the level of the flux in a quarter of the maximum to
the maximum in the rise phase of a flare, the value of b corresponds
to the time in minutes from the maximum to the level of the flux in
a quarter of the maximum in the decay phase of a flare. We determine
the value of the flare index XI as the multiplying the flux in the maximum  
$F_{0.1-0.8}^{max}$ by the
flare duration at the FWQM level that is equal to (a + b). 

$$ XI =~ F_{0.1-0.8}^{max} \cdot (a ~+ b) ~~~~~~~(2) $$

The XI is calculated using the formula (2), the time interval (a+b) is  expressed in seconds. So the XI has the dimension $J/m^2$ and is approximately equal to the total energy calculated as the integral under the flux curve (after
subtraction of the background flux) according to formula (1).

The daily observations of the flux values $F_{0.1-0.8}(t)$ with an
observation interval of 2.5 seconds are available on the GOES
web-site, available at

https://satdat.ngdc.noaa.gov/sem/goes/data/new\_full/ from 2001 to
the present in a real time practically.

To determine the $t_a$ and $t_b$ moments for the flare, according to
the GOES data, we find the moment of maximum $t_m$. By the magnitude
of the flux at the maximum $F_{0.1-0.8}^{max}$, we determine the
level FWQM and find the moments $t_a$ and $t_b$. In Figure 3 $t_a  =
23^h 20^m$, $t_m = 23^h 23^m$ and $t_b  = 23^h34^m$. Accordingly,
the quantity $(a + b) = t^b - t^a = 14$ minutes (840 seconds).

Thus, from the calculations using formula (2) for the flare of
09.03.2011, the X-ray flare index $XI = 1.6E^{-4} \cdot 840 = 0.1344
~J/m^2$. That is, the flare of 09.03.2011 has an X-ray index
XI1.344E-1. For comparison, the total energy $E_{0.1-0.8}$ that is
equal to the area under the flare light curve with allowance for the
background level, calculated according to formula (1), $E_{0.1-0.8} =
0.107 ~J/m^2$.

For 12.07.2012 flare X-ray XI-index  is equal to $XI = 1.4E^{-4}
\cdot 6480 = 0.907 ~J/m^2$. Flare index XI9.07E-1, and, moreover,
the energy), $E_{0.1-0.8} = 0.792 ~J/m^2$. Thus, the value of XI
can be successfully used as a preliminary estimate of the total
energy  $E_{0.1-0.8}$, that in turn is the most important
geoeffective characteristic of the flare (Bruevich \& Bruevich 2018; Reames 2004).

For a comfortable representation of the characteristics of flares,
all information about the
flare can be represented by analogy with
the X-ray classification: the 09.03.2011 flare (Fig.3) of class X1.6
is characterised by the X-ray flare index XI1.3E-1, the 12.07.2012
flare of class X1.4 is characterised by X-ray flare index XI9.1E-1.

A similar representation of XI value as 1.3E-1 and 7.9E-1 is
accepted for use in many computer applications (EXCELL, OrignPRO,
etc.).

\begin{figure}[tbh!]
\centerline{
\includegraphics[width=115mm]{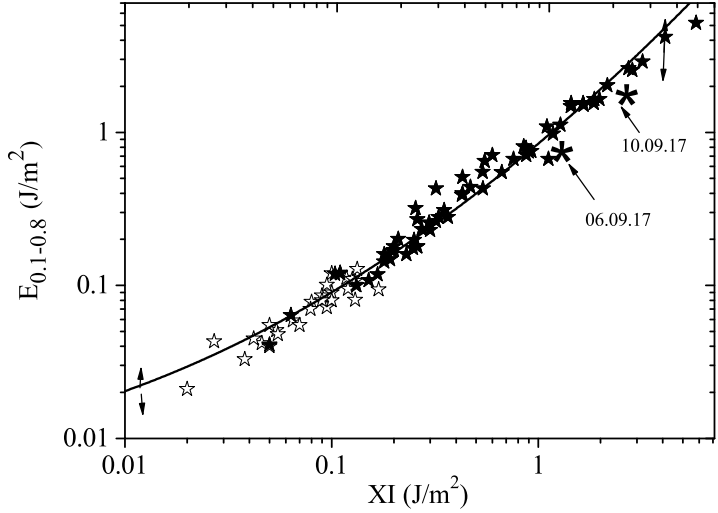}}
 \caption{Dependence of the total flare energy in SXR-range on the new X-ray flare
index XI for 96 flares of cycle 24. Flares with subsequent proton
events are represented by filled asterisks, flares not accompanied
by proton events are represented by hollow asterisks. The quadratic
regression line and the standard deviation are shown. The largest
flares of cycle 24 of 06.09.2017 and 10.09.2017 are marked.}
{\label{Fi:Fig5}}
\end{figure}

If we mean a flare of classes M and C, then it turns out that the XI
values will be two or three orders of magnitude lower than for
flares of class X, because for the classes M and C the magnitude
of the maximum of flares is 1 -- 2 orders of magnitude lower, than
for flares of class X, and because weaker flares have shorter
duration.

That is, for flares of classes M1 -- M9, the X-ray index value may
be the order of XI1E-4 -- XI1E-2 depending on the flare duration,
for the flares of classes C1 -- C4  the XI value varies in the range
of XI1E-5 -- XI1E-3.

Table 1 presents data on 96 major flares in 1998 -- 2017
including data on two largest flares of the 24th cycle that 
occurred in September 2017. We present all the flares of 24th cycle
that are more powerful than M5 and some flares of M2 -- M5 classes
that are characterised by such attendant phenomena such as white light
flares. We also added 21 largest flares of the 23rd cycle.

The information on the X-ray class of flares and their duration at
the FWQM level for a calculation of the X-ray flare index XI according to formula (2) is obtained from the archival data on the GOES
web-site, available at http://www.n3kl.org/sun/noaa\_archive/.

Information on the parameters of flares in the $H_{\alpha}$ line is
available at //www.ngd.noaa.gov/stp/space-weather/solar-data/.

The energy $E_{0.1-0.8}$ of the flares, and the X-ray flare indices
XI are calculated in this paper using equations (1) and (2) and
using the data of the GOES web-site, available at

https://satdat.ngdc.noaa.gov/sem/goes/data / new\_full /.

The relationship between $E_{0.1-0.8}$ and XI is described by the equation:

$$ E_{0.1-0.8}  =~ 0.0198 + 0.658 \cdot XI + 0.106 \cdot XI^2~~~~~~~(3) $$

Figure 5 demonstrates that for a given set of 24th cycle flares,
the X-ray index XI is closely related to the total flare energy in SXR-range 
$E_{0.1-0.8}$, where the value of XI that is calculated according to formula (2) coincides with $E_{0.1-0.8}$ in dimension ($J/m^2$) and
practically coincides in magnitude.

Figure 5 also shows the two largest flares that occurred in
September 2017. It can be seen that the September 6, 2017 X9.3 flare
(the fifth largest since the introduction of the X-ray
classification of flares) significantly loses to the 10.09.2017
flare of X8.2 class as by the value of the total energy in SXR-range 
$E_{0.1-0.8}$, and by the value of the X-ray flare index XI.

\begin{figure}[tbh!]
\centerline{
\includegraphics[width=115mm]{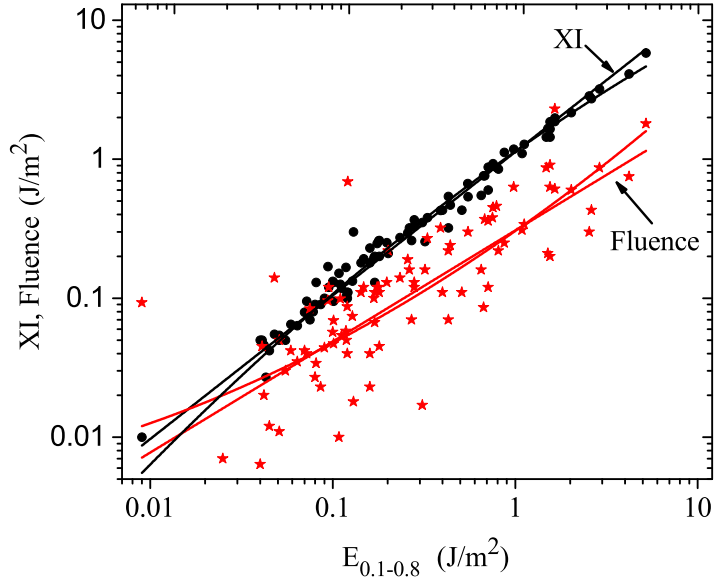}}
 \caption{Dependence of the
index XI and fluence L on the total flare energy in SXR-range $E_{0.1-0.8}$  for 96 flares of cycles 23 -- 24.}
{\label{Fi:Fig6}}
\end{figure}

So-called fluences (L -- the integrated flux from start-max to 1/2 max, in $J/m^{2}$) are often used to estimate the flare energy in the SXR-range (The Catalogue 2008; Preliminary Current Catalogue 2018).

In Figure 6, we can see the dependencies of XI and L versus $E_{0.1-0.8}$. 
Figure 6 shows both linear regression and quadratic regression dependences. It is seen that the dependence of XI vs $E_{0.1-0.8}$ has a smaller scatter of points than the dependence of L vs $E_{0.1-0.8}$.

As a consequence of this fact -- Pearson's linear correlation coefficients for the XI vs $E_{0.1-0.8}$ dependence R=0.93, and for the L vs $E_{0.1-0.8}$ dependence R=0.75. For quadratic regressions, the sum of the residuals RSS is an order of magnitude less in the XI vs  $E_{0.1-0.8}$ dependence than in the L vs $E_{0.1-0.8}$ dependence, which also indicates a closer relationship between the XI-index and the $E_{0.1-0.8}$.

\section{XI-index and SXR-flare flux versus Solar Proton Events (SPEs), CME linear speed and  Kp-index of geomagnetic activity}
{\label{S:place}}

Patrolling observations of the Sun in the SXR-range is the basis of
the modern classification of solar proton events (SPEs), that are
associated with an outstanding flux of high-energy protons.  Such
proton flux is formed as a result of CME which  
accompanes  large solar flares.
It was shown that main role in
acceleration of SEPs play interplanetary CMEs that are the main
source of interplanetary shock waves (Tylka {\it{et al.}} 2005; Reames 2004). In turn, the
coronal CMEs are closely related to large flares originating from
the same active area in the Sun (Klein 2005).
The most significant results on the study of CMEs in recent years were obtained from observations by the LASCO coronagraph on the SOHO spacecraft (see, for example, the reviews by Gopalswamy (2004), Aschwanden (2005), Gopalswamy {\it{et al.}} (2009), Gopalswamy (2016)). The close relationships between flares and CMEs is confirmed by different facts: a similar evolutionary process and explosive energy release in both cases indicates the similarity of temporal velocity profiles of CMEs and of X-ray flares (Zhang {\it{et al.}} 2001). This fact is confirmed by the the existence of close correlation between the kinetic energy of CME and of X-ray fluence (Gopalswamy {\it{et al.}} 2009).

Thus, X-ray flares, observations of which in real time are the most
available for further prediction of geoeffctinve events, are  important objects for a comprehensive analysis and classification associated
with the energy characteristics of the released energy in solar perturbations of the explosive type.

It should be understood that each flare is individual and not all
flare events are manifested in the same way. X-ray observations on
the spacecrafts YOHKOH and RHESSI showed the appearance of centres
of radiation arising from the flare (Podgorny {\it{et al.}} 2009). 
In X-ray photos of
limbic flares, three radiation sources are observed, two of which
are located in the photosphere at the foot of the flare loops. They
are associated with electron beams falling along the field lines,
accelerated in longitudinal currents in accordance with the
prediction of the electrodynamic model. The third source is located
above the flare loop in the corona, where, according to the
electrodynamic model, the radiation source that appears due to the
heating of the plasma after reconnecting in the current layer should
be located (Podgorny {\it{et al.}} 2009).

In Bazilevskaya {\it{et al.}} (2015); Belov {\it{et al.}} (2005) it is shown that most of the favorably located (on the Western part of the solar disk) flares of M5 class are more powerful are accompanied by SPEs, and almost all
SPEs can be identified with a particular solar flare. The duration
and intensity of the injection of protons for the observer in the
Ecliptic plane varies from one proton event to another proton event by
several orders of magnitude.

The physical basis for the communication of SPEs with SXR-radiation is the fact that the source of heating of the flare plasma
can be accelerated electrons, that are accelerated simultaneously
with protons. At the same time, there is no theory that connects
quantitatively different types of electromagnetic and corpuscular
radiation of solar flares (Belov {\it{et al.}} 2005). According to statistical analysis
(Li {\it{et al.}} 2013; Sharykin {\it{et al.}} 2012)  that was made for several largest proton events of the
23rd and 24th cycle, the time difference from the maximum flux $F_{0.1-0.8}^{max}$ of the flare to the maximum flux of SPE is approximately equal to 35 minutes -- 2 hours depending on the path of proton propagation.

In this case, from the graphic data GOES
(www.n3kl.org/sun/noaa\_archive/) it can be seen that the phase of a
sharp increase in the proton flux coincides with the phase of the
flare flux decay. Most often, the proton flux ($I_{pr}$) quickly
reaches a maximum and for some time (depending on the flare power -
up to a couple of days) is kept at a constant level and then slowly
decreases.

An important property of the X-ray index XI, as an energy
characteristic of the flare, is its obvious connection with proton
flares (proton events). Usually, when discussing the relationship
between thermal and non-thermal electromagnetic radiation of solar
flares, we refer to the similarity of time profiles of the intensity
of non-thermal radiation and the derivative on time of SXR-flux (Neupert effect). This similarity corresponds to a single-loop
evaporation model, but it is not observed in more than 50\% of
long-term events and this is due to the long and multiple
acceleration of electrons with a variable spectrum in the system of
flare loops in different physical conditions (Struminskii 2011).

In the study of the energy release of solar flares, it is necessary
to have an idea of the fine structure of the flare region, since
many energy release channels depend on the geometric parameters of
the magnetic loops along which the energy transfer occurs. The fine
structure can affect the density of the accelerated electrons in the
beam and their propagation in the plasma.

All these features of various flares manifest themselves in
the ambiguity of the relationship between the XI-index, X-ray fluxes in flares and SPEs, as evidenced by the large scatter of values in Figures 7 and 8.

According to the forecast center RUSSIAN HELIOGEOPHYSICAL MONITORING CENTER (http://space-weather.ru/index.php?page=home-en) a geoeffective flare (an event associated with CME) is an event that entails the following change in geomagnetic activity indices:$ D_{st}  \leqslant $ - 80 and Kp $\geqslant$  7. With such values of geomagnetic indices, magnetic storms become dangerous both for aviation when flying at an altitude of more than 10,000 m, and ground-based radio communication devices, etc. Also, the onset of GLE is undoubtedly a geoeffective event in which$ D_{st}$ can be less than -250-300, and Kp $\geqslant$  8.

Traditionally, a statistical analysis of flares with subsequent SPE
is carried out to identify patterns of interaction between X-ray
bursts and geoeffective proton events. Since 1970, SPEs, in which the
protons with energy E $\geqslant$ 10 MeV and fluxes $I_{pr} \geqslant 1$ pfu
(1 pfu = 1 $cm^{-2} \cdot s^{-1} \cdot sterad^{-1}$) were observed,
has been collected in catalogs edited by Yu. I. Logachev, available
at www.wdcb.ru/stp/online\_data.ru.html. These catalogs are
characterized by homogeneous and long time series.

According to flares  of 23rd and 24th cycles,  only the most common
patterns of parent flares and proton fluxes were revealed in Bazilevskaya {\it{et al.}} 2015:
when the maximum flare  amplitude in the SXR-range
was changed from $10^{-6} ~$to $~10^{-3}~ W \cdot m^{-2}$, the maximum
proton flux $I_{pr}$ increased from 0.01 to 100 pfu for protons with
E $\geqslant $ 100 MeV. The largest flares that are accompanied by
so-called ground-based increases (GLEs) are among the most powerful
SPEs and proton fluxes with E $\geqslant $ 100 MeV for such flares are
enclosed in the range of 100 -- 1000 pfu.

The flares accompanied by GLE are of great interest both for the
determination of the mechanisms of acceleration and propagation of
charged particles in the Sun and in the interplanetary medium, and
for the determination of radiation danger in the near-Earth space.
According to the catalogs (The Catalogue 2008; Preliminary Current Catalogue 2018), in 24th cycle only three events
with GLE were recorded (17.05.2012, 06.01.2014 and 10.09.2017),
while 15 events with GLE were recorded in the cycle 23.

From The Catalogue (2008) and Preliminary Current Catalogue (2018) it also follows that in 2018, when we are
already at the minimum between 24th and 25th cycles, the number of
events with energy of protons E $ \geqslant $ 10 MeV throughout 24th cycle is
approximately 30\% less than in cycle 23 (60 versus 89,
respectively). The comparison of flare parameters from Table 1 is shown
in Figure 5. Blackened asterisks indicate flares that accompanied by
SPEs with different levels of flux but exceeding the value of 10 pfu
for protons with E $\geqslant $ 10 MeV. 

Figure 5 demonstrates that flares with
subsequent proton event are most likely characterized by X-ray index
greater than XI1.0E-1 -- XI1.5E-1. This paper examines the
relationship between the so-called parent flares and the SPEs that 
follow them.

Table 1 shows all the flares of 23rd and 24th cycle that are
accompanied by SPEs and characterised by $I_{pr}\geqslant 2$ pfu, as well as
the most significant flares of 23rd cycle  (21 flares with $I_{pr}$
that are characterized by proton fluxes of 100 -- 10000 pfu for SPsE
with E $\geqslant $ 10 MeV). Some of these flares are accompanied by
significant SPEs with E $\geqslant $ 100 MeV.

In Figure 7 and in Figure 8, the dependencies of the X-ray index XI and $F_{0.1-0.8}^{max}$ 
vs values of proton fluxes with E $\geqslant $ 10 MeV and E $\geqslant $ 100 MeV  for
96 parent flares from Table 1 are presented. The quadratic regression lines are marked.
Data on proton fluxes in the ranges with E $\geqslant $ 10 MeV and E $\geqslant $ 100
MeV are taken from The Catalogue (2008) and Preliminary Current Catalogue (2018) and also refined directly from
the GOES archive data, available at
(www.n3kl.org/sun/noaa\_archive/).

\begin{figure}[h!!!]
\centerline{\includegraphics[width=110mm, height=75mm]{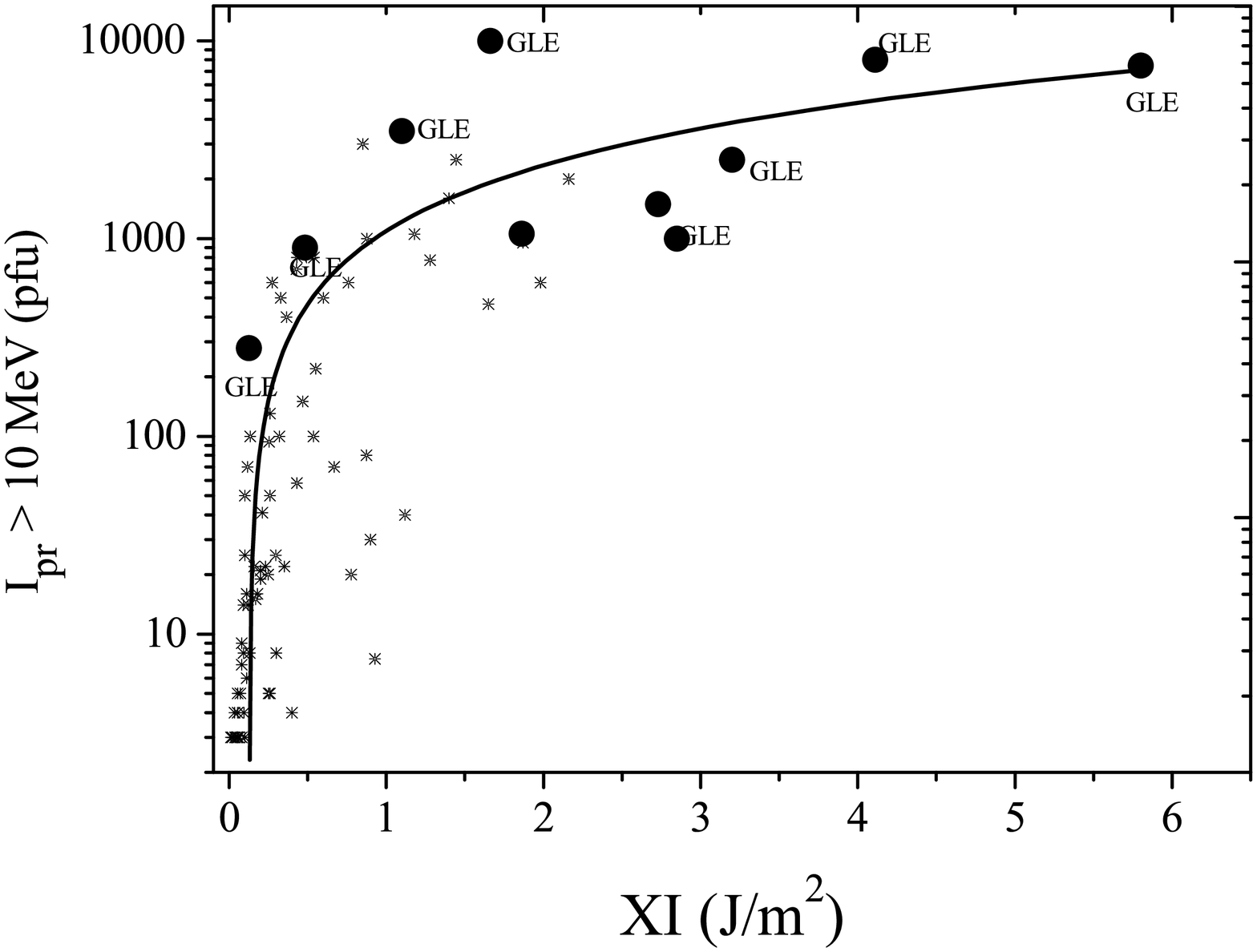}}
\vspace{0.05\textwidth}   
\centerline{\includegraphics[width=110mm, height=75mm]{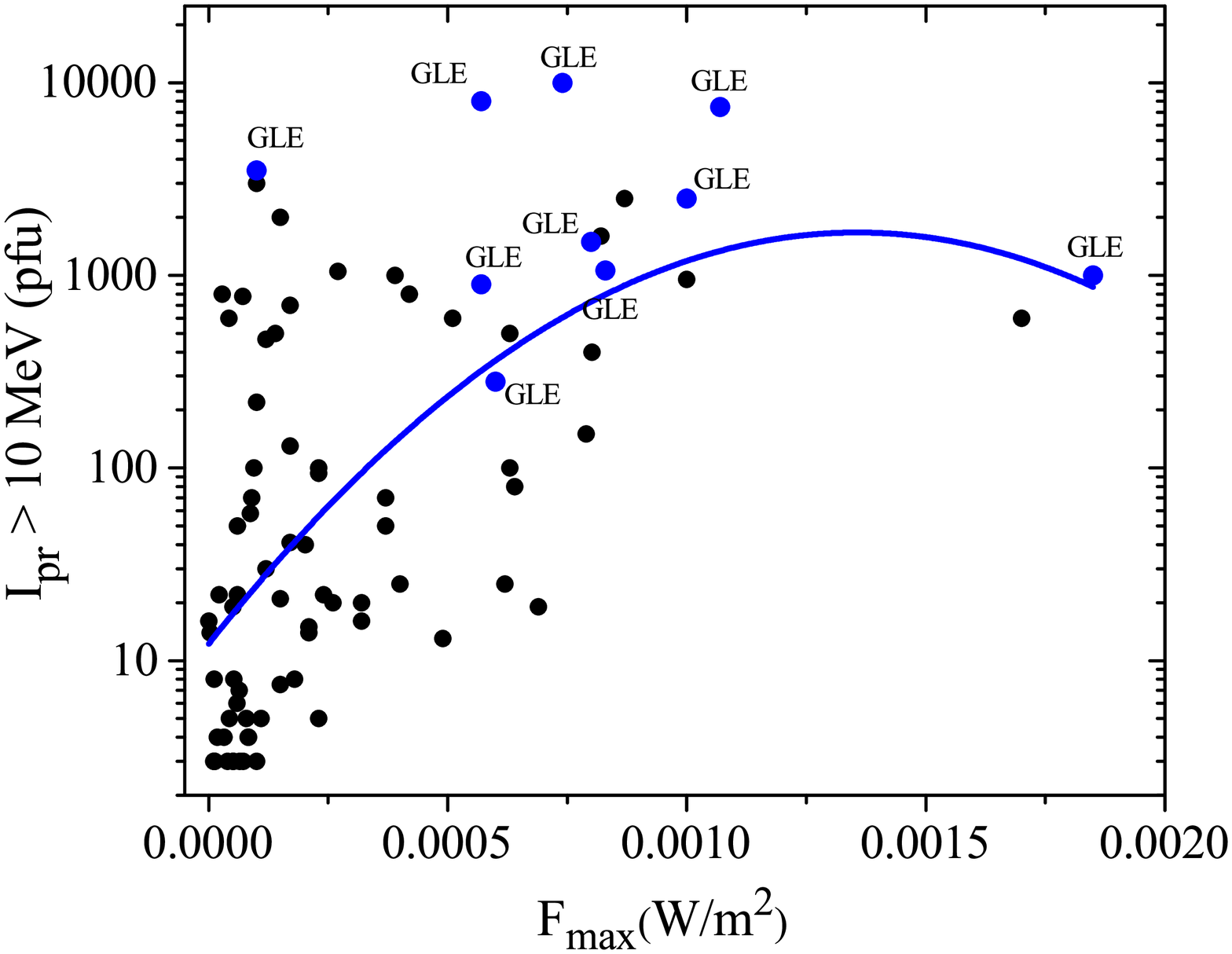}}
\caption{Top: The dependence of the  proton flux with energies E $ \geqslant $ 10 MeV  versus X-ray index XI of the parent flares in cycles 23 and 24;
Bottom: The dependence of the  proton flux with energies E $ \geqslant $ 10 MeV  versus $F_{0.1-0.8}^{max}$. The Ground Level Events (GLE) and quadratic regression lines are marked.}
\label{Fig7}
\end{figure}

\begin{figure}[h!!!]
\centerline{\includegraphics[width=110mm, height=75mm]{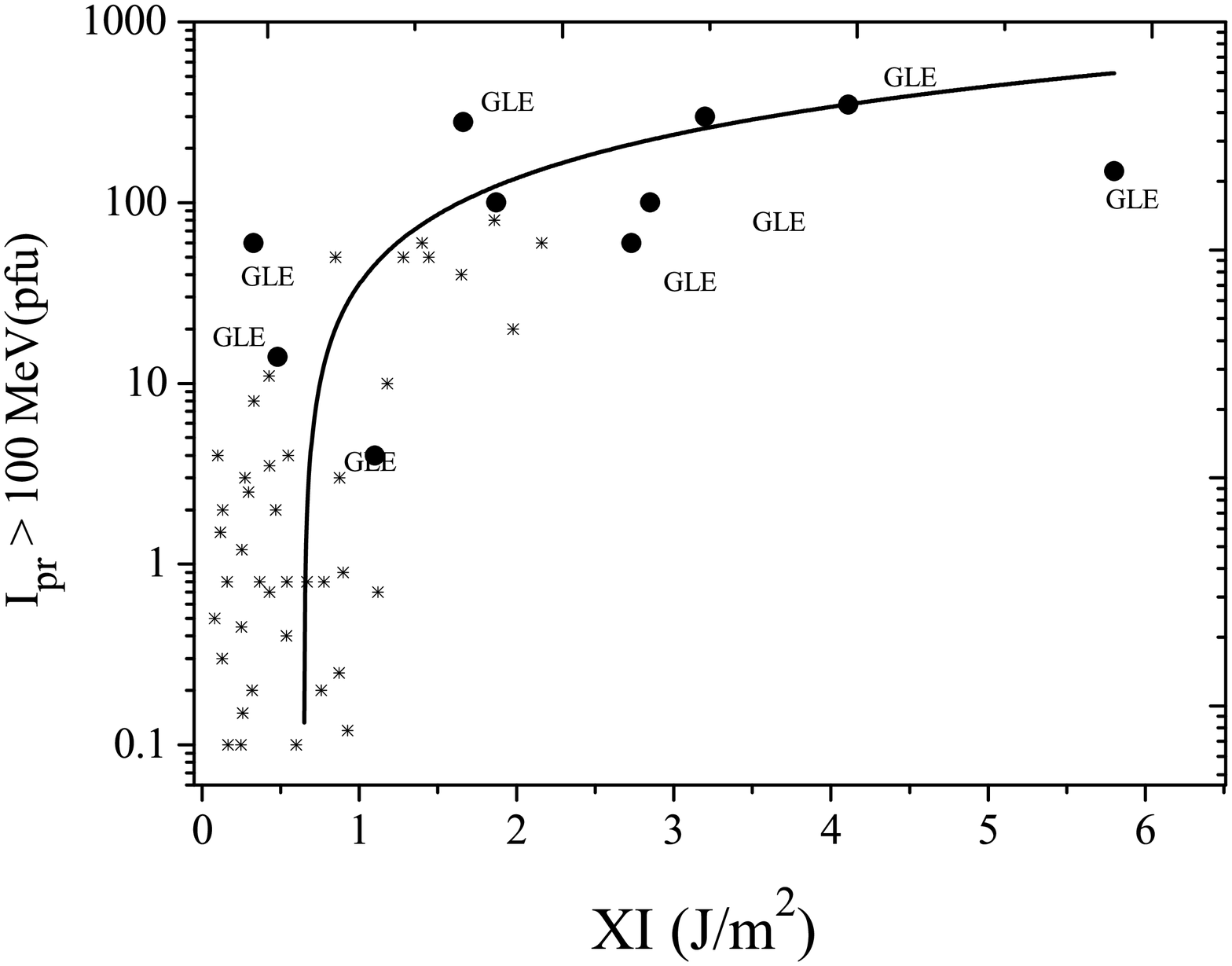}}
\vspace{0.05\textwidth}   
\centerline{\includegraphics[width=110mm, height=75mm]{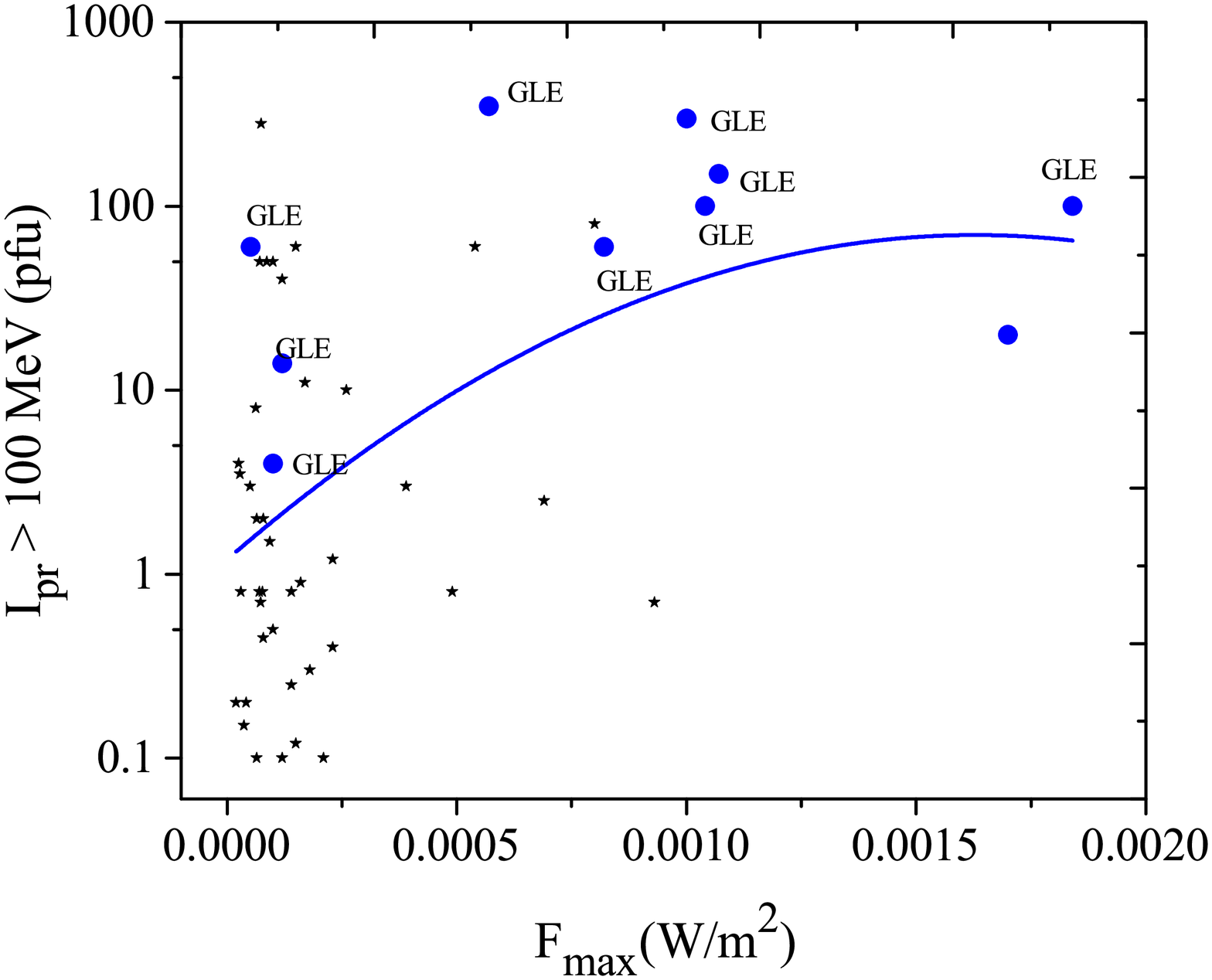}}
\caption{Top: The dependence of the  proton flux with energies E $ \geqslant $ 100 MeV  versus X-ray index XI of the parent flares in cycles 23 and 24;
Bottom: The dependence of the  proton flux with energies E $ \geqslant $ 100 MeV versus $F_{0.1-0.8}^{max}$. The Ground Level Events (GLE) and quadratic regression lines are marked.}
\label{Fig8}
\end{figure}

\begin{figure}[h!!!]
\centerline{\includegraphics[width=110mm, height=75mm]{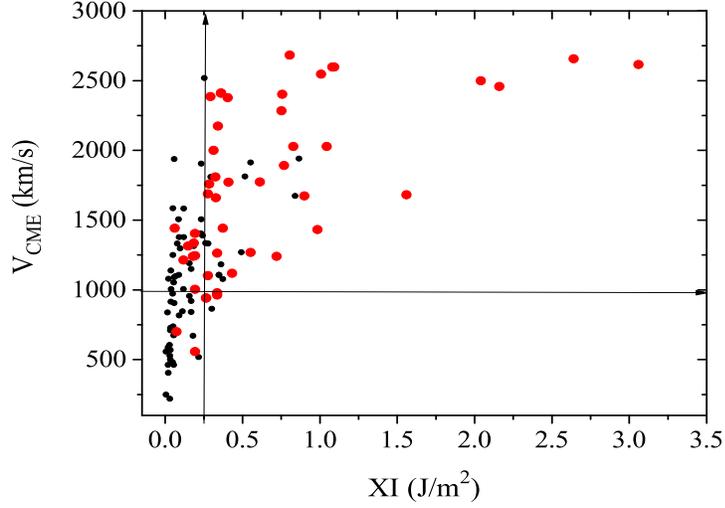}}
\vspace{0.05\textwidth}   
\centerline{\includegraphics[width=110mm, height=75mm]{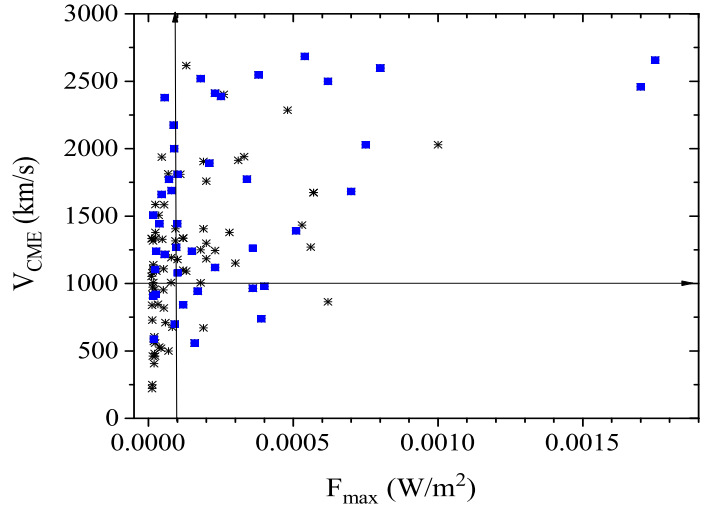}}
\caption{Top: The dependence of CMEs linear speed  $V_{CME}$ on the X-ray index XI in cycles 23 and 24; 
Bottom: the dependence of the CME linear speed $V_{CME}$ versus $F_{0.1-0.8}^{max}$.}
\label{Fig9}
\end{figure}

\begin{figure}[h!!!]
\centerline{\includegraphics[width=110mm, height=75mm]{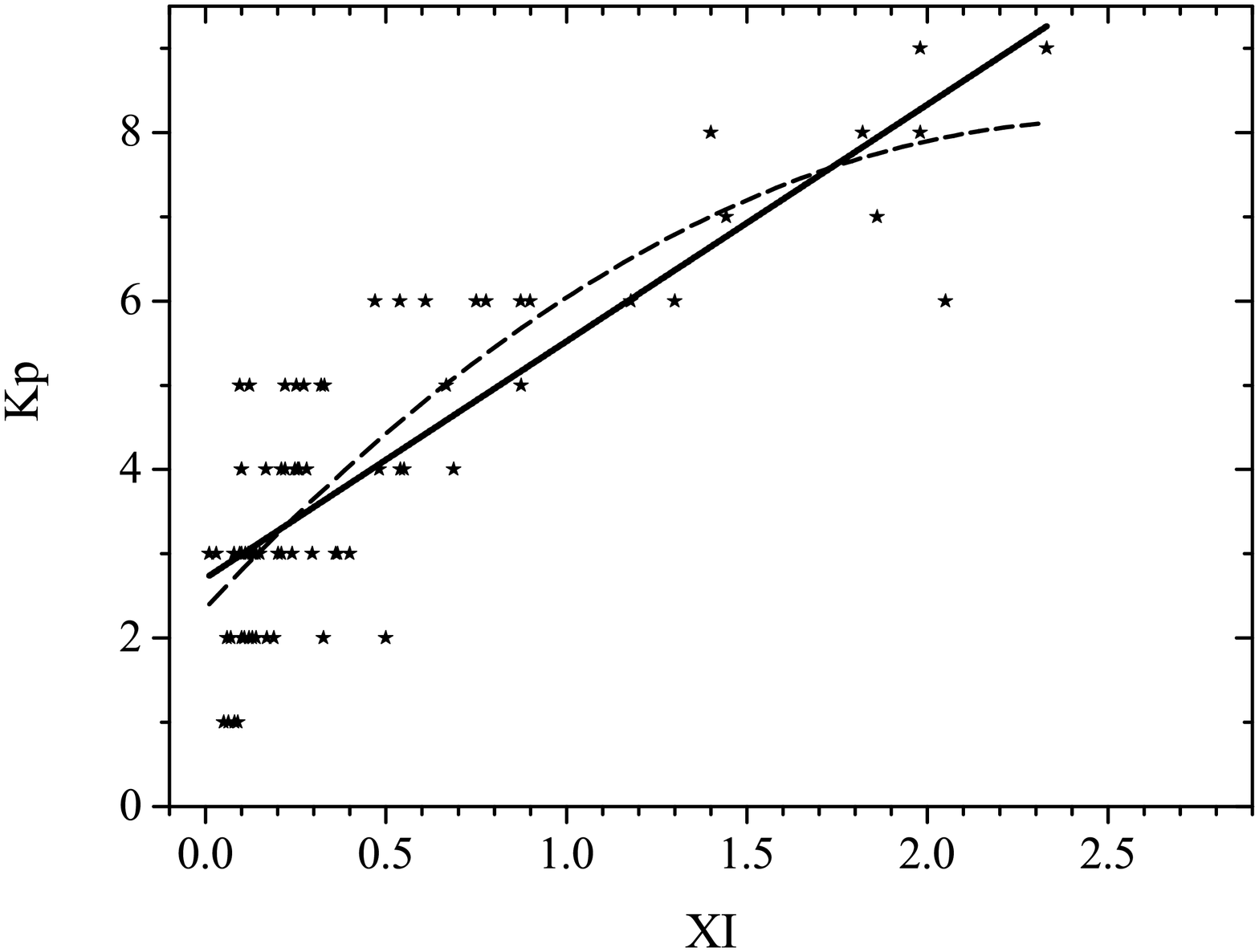}}
\vspace{0.05\textwidth}   
\centerline{\includegraphics[width=110mm, height=75mm]{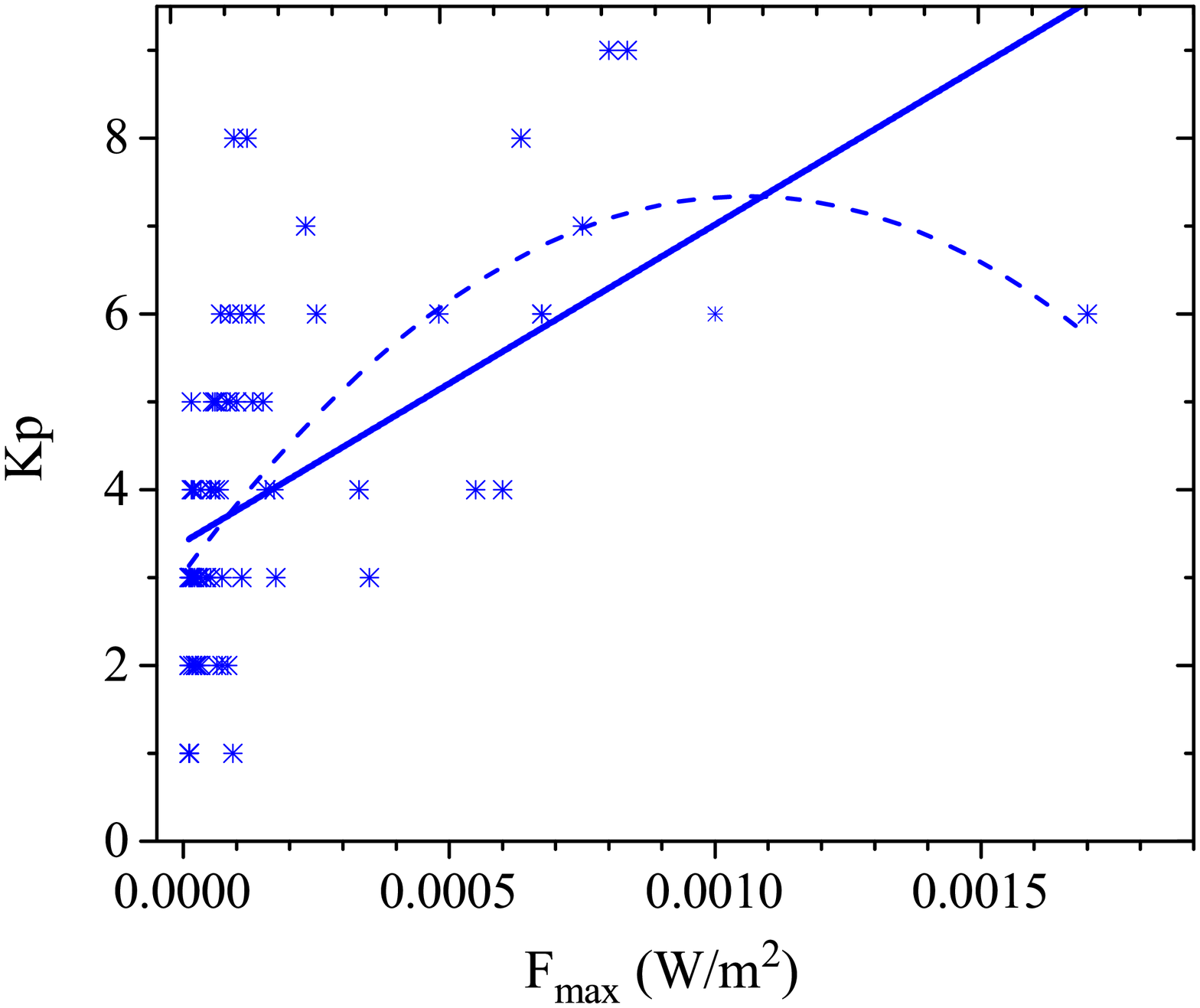}}
\caption{Top: The dependence of the Kp-index of geomagnetic activity on the X-ray index XI in cycles 23 and 24;
Bottom: the dependence of the Kp-index of geomagnetic activity versus $F_{0.1-0.8}^{max}$. Linear and quadratic regression lines are marked.}
\label{Fig10}
\end{figure}

\begin{table}
\caption{Parameters of 96 flares of 23rd and 24th cycles and
subsequent Solar Proton Events affecting the earth environment}
\begin{center}
\begin{tabular}{clclclclclclcl}

\hline
   Date of flare&X-ray class/&X-ray index&Proton flux $I_{pr}$ &Proton flux $I_{pr}$ \\

~~~ &$H_\alpha$-score/&XI &E $ \geqslant $ 10 MeV,&E $\geqslant $ 100 MeV \\

~~~ &$H_\alpha$-duration, min& ($J/m^2$) &~~pfu &~~pfu \\
\hline

24.08.1998& X1.0/3B/180&    0.55&   220&    4 \\
30.09.1998& M2.8/2N/150&    0.43&   800&    3.5 \\
14.07.2000& X5.7/3B/105/GLE&    4.11&   8000&   400& \\
08.11.2000& M7.4/3F/80/GLE& 0.66&   10000&  300& \\
24.11.2000& X2.3/2B/22& 0.255&  94& 1.2 \\
02.04.2001& X18.4/1B/70&    2.85&   1000&   5.5 \\
09.04.2001& M7.9/1B/70&     0.251&  5&  0.45 \\
10.04.2001& X2.3/3N/176&    0.538&  100&    0.4 \\
15.04.2001& X10.4/3B/130/GLE&   1.87&   951&    100.3  \\
24.09.2001& X2.6/2B/155&    1.178&  1050&   10 \\
01.10.2001& M4.2/2B/160&    0.76&   600&    0.2 \\
04.11.2001& X1.0/3B/450/GLE&    0.85&   3000&   50 \\
22.11.2001& X1.0/2B/160/GLE&    1.1&    3500&   4 \\
26.12.2001& M7.1/1B/230/GLE&    1.28&   779&    50 \\
21.04.2002& X1.5/1F/125&    2.16&   2000&   20 \\
26.10.2003& X1.2/1N/170&    1.65&   466&    1.0 \\
28.10.2003& X10.7/4B/280/GLE&   5.8&    7500&   150 \\
29.10.2003& X10.0/2B/150/GLE&   3.2&    2500&   100 \\
02.11.2003& X8.3/2B/170&    1.86&   1060&   40 \\
03.11.2003& X3.9/2F/50&     0.875&  1000&   3 \\
04.11.2003& X17/3B/80&  1.98&   600&    1.1 \\
28.01.2011& М1.3/1F/30&    0.01&   3&  - \\
15.02.2011& X2.2/2F/25& 0.02&   3&  - \\
07.03.2011& M3.7/1F/90& 0.259&  50& 0.15 \\
07.06.2011& M2.5/2N/100&    0.1&    50& 4 \\
14.06.2011& M1.3/SF/40& 0.30&   8&  - \\
04.08.2011& M9.3/2B/84& 0.118&  70& 1.5 \\
08.08.2011& M3.5/1B/55& 0.054&  4&  - \\
09.08.2011& X6.9/2B/72&     0.296&  25& 2.5 \\
06.09.2011& X2.1/2B/55&     0.167&  15& 0.1 \\
07.09.2011& X1.8/3B/74& 0.13&   8&  0.3 \\

\hline

\end{tabular}
\end{center}
\end{table}

\begin{table}
\caption{Continuation of Table 1}
\begin{center}
\begin{tabular}{clclclclclclcl}

\hline
   Date of flare&X-ray class/&X-ray index&Proton flux $I_{pr}$ &Proton flux $I_{pr}$ \\

~~~ &$H_\alpha$-score/&XI &E $\geqslant $ 10 MeV,&E $\geqslant $ 100 MeV \\

~~~ &$H_\alpha$-duration, min& ($J/m^2$)  &~~pfu &~~pfu \\

\hline
22.10.2011& M1.3/1N/130 &   0.26&   5 & - \\
03.11.2011& X1.9/2B/82&      0.4&   4 & - \\
25.12.2011& M4.0/1N/60& 0.054 & 3.0 & - \\
23.01.2012& M8.7/2B/335&    1.443&  2500&   2.2 \\
27.01.2012& X1.7/2F/96&     0.428&  700&    11 \\
05.03.2012& X1.0/2B/105&    0.1&    3&  - \\
07.03.2012& X5.4/3B/220&    1.40&   1600&   60 \\
09.03.2012& M6.3/SF/160&    0.329&  500&    8 \\
13.03.2012& M7.9/1B/185&    0.469&  150&    2 \\
17.05.2012& M5.1/1F/93/GLE& 0.126&  280&    20 \\
14.06.2012& M2.1/2B/150&    0.12&   14& - \\
06.07.2012& M6.2/1B/40& 0.1&    25& - \\
08.07.2012& M6.9/1N/23& 0.2&    19& - \\
12.07.2012& X1.4/2B/250&    0.873&  80& 0.25 \\
17.07.2012& M1.7/1F/250&    0.26&   130&    - \\
19.07.2012& M7.7/SF/155&    0.667&  70& 0.8 \\
08.11.2012& M1.7/1F/40& 0.05&   3&  - \\
14.11.2012& M1.1/1F/10& 0.08&   9&  - \\
15.03.2013& M1.1/1N/180&    0.18&   16& - \\
11.04.2013& M6.5/3B/125&    0.133&  100&    2 \\
13.05.2013& X1.7/1N/39& 0.21&   41& - \\
15.05.2013& X1.2/2N/66&     0.248&  20& 0.1 \\
22.05.2013& M5.0/3N/180&    0.273&  600&    3 \\
21.06.2013& M2.9/1F/87& 0.11&   6&  - \\
23.06.2013& M2.9/1N/12& 0.09&   14& - \\
28.10.2013& X1.0/2N/50& 0.07&   5&  - \\
29.10.2013& X2.3/1N/20& 0.055&  5&  - \\
01.11.2013& M6.3/1B/70& 0.06&   3&  - \\
06.11.2013& M1.8/1F/40& 0.08&   7&  - \\
19.11.2013& X1.0/SF/80& 0.035&  4&  - \\

\hline

\end{tabular}
\end{center}
\end{table}

\begin{table}
\caption{Continuation of Table 1}
\begin{center}
\begin{tabular}{clclclclclclcl}

\hline
   Date of flare&X-ray class/&X-ray index&Proton flux $I_{pr}$ &Proton flux $I_{pr}$ \\

~~~ &$H_\alpha$-score/&XI &E $\geqslant $ 10 MeV,&E $\geqslant $ 100 MeV \\

~~~ &$H_\alpha$-duration, min& ($J/m^2$)  &~~pfu &~~pfu \\

\hline

07.01.2014& X1.2/2N/76/GLE &    0.481&  900&    4 \\
20.02.2014& M3.0/SN/60& 0.16  & 22& 0.8 \\
25.02.2014& X4.9/2B/90&     0.777&  20& 0.8 \\
29.03.2014& X1.0/2B/40&     0.0792& 3&  0.5 \\
18.04.2014& M7.3/1N/50& 0.43&   58& 0.7 \\
10.09.2014& X1.6/2B/240&    0.899&  30& 0.9 \\
13.12.2014& M1.5/1F/15& 0.05&   3&  - \\
20.12.2014& X1.8/3B/90& 0.038&  3&  - \\
15.03.2015& M1.2/1F/40& 0.09&   8&  - \\
18.06.2015& M1.2/1N/80& 0.11&   16& - \\
21.06.2015& M2.0/1N/120&    0.32&   100&    0.2 \\
21.06.2015& M6.5/2B/170&    0.6&    500&    0.1 \\
25.06.2015& M7.9/3B/63&0.35&22&- \\
20.09.2015& M2.1/2N/120&    0.05&   3&  - \\
09.11.2015& M3.9/2N/65&     0.095&  4&  - \\
28.12.2015& M1.8/1F/120&    0.02&   3&  - \\
01.01.2016& M2.3/1N/105&    0.2&    21& - \\
14.07.2017& M2.4/1N/180&    0.23&   22& - \\
04.09.2017& M7.0/2N/90&     0.54&   800&    0.8 \\
06.09.2017& X9.3/2B/40&     1.12&   40& 0.7 \\
07.09.2017& X1.4/2B/120&    0.366&  400&    0.8 \\
10.09.2017& X8.2/3B/120/GLE &   2.73&   1490&   60 \\

\hline

\end{tabular}
\end{center}
\end{table}

Figure 7 and Figure 8 show that the spread in values of proton
fluxes from quadratic regression lines (that are marked in the figures) is sufficiently large, which confirms the complexity and
variety of flares and associated SPEs. 
It can also be seen that in cases when SPEs of different power are analysed depending on the XI- index as compared to the dependence on the $F_{0.1-0.8}^{max}$, the spread of values in the first case (XI- index) is less. Note that   when we use the linear (not log) scales in Figures 7a,8a and 7b,8b and analise the dependencies with linear regressions, then for dependencies in Figures 7a,8a the Pearson's correlation coefficient will be approximately equal to 0.6-0.7, and for Figures 7b,8b -- approximately 0.3-0.4.
We can make an assessment (according to Figure 7a) that in case when XI$>$ 0.5 and Ipr$>$ 200 are simultaneously satisfied, then the probability of GLE is approximately 0.5.

In Figures 9a, 9b the possibility of the simple prediction of the geoeffective event characterised by high geomagnetic activity $D_{ST} \leqslant  -80$ caused by explosive events in the Sun demonstrate. As noted above, the similarity of CME speed $V_{SME}$ profiles to that of flare soft X-ray emission peculiar properties takes place (Zhang {\it{et al.}} 2001). 

Figure 9a shows the relationship between the $V_{SME}$ and the XI-index  for the flares with CMEs, taking into account the dependence on the value of the $D_{ST}$ index. Large circles  correspond to the case with geomagnetic index $D_{ST} \leqslant  -80$. In Figure 9b -- the relationship between the $V_{SME}$ and the $F_{0.1-0.8}^{max}$ for the flares with CMEs, taking into account the dependence on the value of the $D_{ST}$ index. Large squares  correspond to the case with geomagnetic index $D_{ST} \leqslant -80$.

In Figure 9a it was shown a region characterised by $V_{SME} \geqslant 1000$ together with XI $\geqslant 0.25$. In this region, there are 46 points, among which 35 points are geoeffective  with $D_{ST} \leqslant   -80$ (large circles). Thus, the probability of a geoeffective event with $D_{ST} \leqslant  -80$ is 35/46=0.76. 

In Figure 9b, it was shown a region characterized by $V_{SME} \geqslant  1000$ together with $F_{0.1-0.8}^{max} \geqslant 0.0001 $ (Flares of class more or equal to X1). The $F_{0.1-0.8}^{max}$ values on Figure 9b were chosen so that the number of points in the selected area coincided with the number points in the selected area in Figure 9a and was also equal to 46. The probability of a geoeffective event with $D_{ST} \leqslant -80$ in this case is 26/46=0.56.
The advantage of the XI-index is obvious in this case.

Figure 10 shows the dependencies of the global planetary Kp-index of geomagnetic activity  on the XI-index of flares, which, together with the associated CMEs, are the sources of magnetic storms. The Kp-index ranges from 0 to 9, where a value of 0 means no geomagnetic activity, and a value of 9 means an extreme geomagnetic storm. For the events of cycles 23 and 24 from our sample, there were no magnetic storms with Kp = 9.
The flares (or their XI-indexes) on the  Figures 10a,b are the same as those presented in Table 1. Note that not all flares from Table 1 were accompanied by geomagnetic activity (SPEs, magnetic storms), since events that subsequently become geoeffective should come from active regions located in the western part of solar disk.

It can be seen that the interconnection between the variables in the case of Figure 10a is closer than in the case of Figure 10b (the Pearson's correlation coefficient R=0.75 in Figure 10a vs. R=0.45 in Figure 10b).

\begin{figure}[tbh!]
\centerline{
\includegraphics[width=115mm]{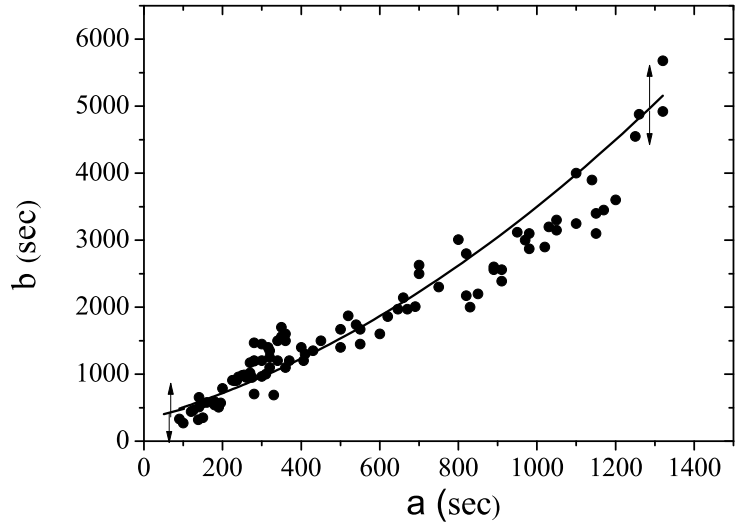}}
 \caption{The dependency of a from b for 96 flares of cycles 23 and 24 studied in this work.
 The quadratic regression line and the standard deviation are shown.}
{\label{Fi:Fig9}}
\end{figure}

\begin{figure}[tbh!]
\centerline{
\includegraphics[width=115mm]{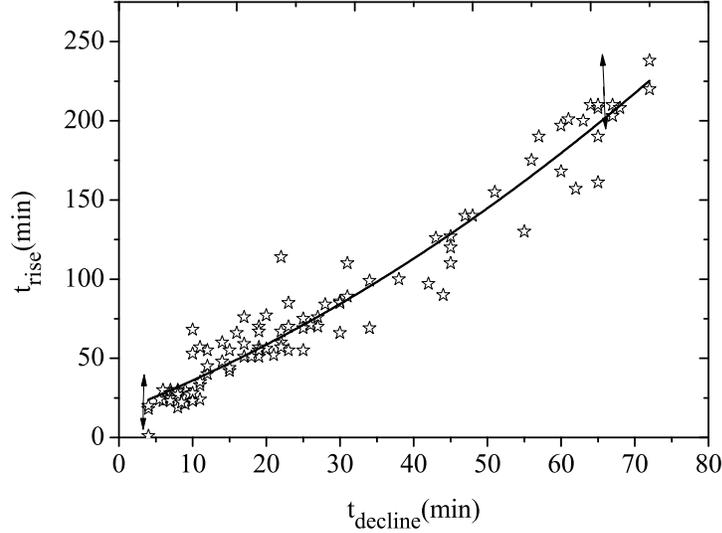}}
 \caption{The dependency of $t_{rise}$ from $t_{desay}$  for 96 flares of cycles 23 and 24
 studied in this work. The quadratic regression line and the standard deviation are shown.}
{\label{Fi:Fig10}}
\end{figure}

Note that the calculation of
the X-ray index XI for a particular flare is very simple: we have to
determine from the GOES archive data the level of FWQM that is equal
to one quarter of the flux at the maximum, then to determine the
moments $t_a$ and $t_b$ and to use formula (2).

Observations of GOES are presented with an interval of 2.5 seconds, 
which entails an error in determining of the X-ray index XI only in
the third-fourth significant digit (less than 0.1\%). A calculation
the value of energy $E_{0.1-0.8}$ (as the flare characteristics that
is compared to XI) requires much more efforts: to integrate the
flare light curve, you need to convert the archived GOES data,
linking them to the real time-scales and correctly select the
background level. Finally, the error in the definition of
$E_{0.1-0.8}$ can already reach several percent. In particular, due
to the fact that the background level before and after the flare may
vary, and this introduces uncertainty in determining the flare end
time.

\section{Relationship between rising and declining phases of solar flares}
{\label{S:place}}

To calculate the index XI, it is important to determine the times
$t_a$ and $t_b$ (intervals a and b) most precisely. For GOES data,
this is not difficult, except when flares go on overlapping one
another before they reach the level of the background flux during
the phase of decline. In this case, the following comparative
analysis of the data on the relationship between the phase of rise
and phase of decline of flares can help.

In this paper, we investigated the relationship between the
parameters $a$ and $b$ for a sample of 96 significantly more
powerful flares of classes M2 -- X9 from the Table 1.

The relationship between the parameters $a$ and $b$, expressed in
seconds, is shown in Figure 9. The relationship between the values
of $a$ and $b$ is described by the quadratic regression equation:

$$ b = 356 + 1.32 \cdot a + 0.0018 \cdot a^2        ~~~~                        (4)$$

Thus, taking into account the dependence (4), it is possible to
estimate the time $t_b$ in complex cases, when flares follow one
after another and one flare is applied to subsequent flare. For
some flares, there is a discrepancy in the determination of the time
of beginning and ending of the flare in $H_\alpha$ in different
observatories. Sometimes observers simply indicate that the flare
lasts more than a certain time. This introduces uncertainty in the
calculation of the FI flare index.

To correctly determine the duration of $H_\alpha$ flares for 96 flares from Table 1, the connection between the
duration from the beginning of the flare to its maximum ($t_{rise}$)
and the duration from the flare maximum to its end ($t_{decay}$) is
studied. $t_{rise}$ and $t_{decay}$ correspond to a full interval of
time in the rise phase and in the decline phase of the flare, taking
into account the excess of radiation from flares above the level of
background in $H_\alpha$.

In Figure 12, the relationship between $t_{rise}$  and $t_{decay}$ is described
by a quadratic regression equation with a relatively small
second-order term:

$$ t_{decay} = 17.68 + 2.006  \cdot  t_{rise} + 0.0145 \cdot  t_{rise}^2      ~~~~                    (5)$$

Using dependencies (4) and (5), you can determine the values of the
time intervals $b$ and $t_{decay}$ that are necessary to calculate
the X-ray index XI and the optical index FI in cases where it is
impossible to determine the parameters $b$ and $t_{decay}$ directly
from observations.

\vskip12pt
\section{Conclusion}
\vskip12pt

The method of additional flare classification proposed in this paper based on determining the X-ray flare index XI by analogy with the optical flare index FI has the
following advantages:

1.  The X-ray index XI is easily calculated according to formula (2). Data on
the values of $a$, $b$, $F_{0.1-0.8}^{max}$, $F_{background}$ are
available on the GOES web-site (from 1978 to the present).
Accordingly, for each flare, the X-ray index XI can be calculated,
starting from 1978.

2. X-ray index XI is an analog of the total energy of flare in SXR-range 
$E_{0.1-0.8}$ that can be calculated according to formula (1). The relationship between
XI and $E_{0.1-0.8}$ is described by (3), resulting if you 
know the index XI, you can rapidly evaluate the flare parameter $E_{0.1-0.8}$.

3. By the value of the index XI, as well as by the value
$E_{0.1-0.8}$, it is possible to determine flares with subsequent
SPEs (under the condition of localization of the flare
region in the western part of the Sun's disk suitable for
propagation of protons towards the Earth).

4. X-ray index XI, as well as $E_{0.1-0.8}$, is the most important
geoeffective  parameter of the flare. The combined use of XI and of the most important parameters of flares, as well as associated CMEs, it is possible to predict the probability of occurrence of geoeffective events of various scales, that characterised by the geomagnetic indices Dst and Kp.

\vskip12pt
\vskip12pt
\begin{Large}
\centerline{
{\bf{References}}
}
\end{Large}
\vskip12pt
\vskip12pt

\begin{flushleft}

Altyntsev A. T., Banin V. G., Kuklin G. V., Tomozov V. V. 1982, 
Solar Flares, Moscow.: Nauka

Aschwanden M. J. 2005, Physics of the Solar Corona. Springer, Berlin.

Bazilevskaya G. A., Logachev Yu. I., Vashenyuk E.V. {\it{et al.}} 2015, 
Bulletin of the Russian Academy of Sciences: Physics, 79, p. 627

Belov A., Garcia N., Kurt V. {\it{et al.}} 2005, Solar Physics, 229, p. 135

Bogod V. M. 2006, Bulletin of the Russian Academy of Sciences: Physics, 
70, p. 491

Bogod V. M. 2011, Astrophysical Bulletin, 66, p. 190

Bowen T. A., Testa P., Reeves K. K. 2013a, LWS/SDO Science Workshop, 
(SOC), Cambridge, USA

Bowen T. A., Testa P., Reeves K. K. 2013b, Astrophysical Journal, 770, 
p. 126 

Bruevich E. A., Yakunina G. V. 2017, Astrophysics, 60, p. 387

Bruevich E. A., Bruevich V. V. 2018, Astrophysics, 61, p. 241

Gopalswamy N. A. 2004,  Astrophys. Space Sci. Libr., 317. p. 201

Gopalswamy N., Yashiro S., Michalek G., Stenborg G., Vourlidas A.,  Freeland S., Howard R. 2009, Earth, Moon, and Planets, 104, p. 295

Gopalswamy N. 2016, Geoscience Letters, 3, article id.8, 18 pp

Kleczek J. 1952, Publ. Czech Centr. Astron. Inst., 22, p. 1

Klein K.-L., Krucker  S., Trottet G., Hoang S. 2005, Astron. Astrophys., 431, p. 1047

Li C., Firoz K. A., Sun L. P., Miroshnichenko L. I. 2013, ApJ, 770, article id. 34, 12 pp

Ozgus A., Atac T., Rybak J. 2003, Solar Physics, 214, p. 375

Preliminary Current Catalogue of Solar Flare Events. 2018, 

//www.wdcb.ru/stp/data/Solar\_Flare\_Events/Fl\_XXIV.pdf

Podgorny I. M., Vashenyuk E. V., Podgorny A. I. 2009, Geomagnetism and Aeronomy, 
49, p. 1115

Reames D. V. 2004, Adv. Space Res., 34(2), p. 381

Sharykin I. N., Struminsky A. B., Zimovetz I. V. 2012, Astronomy Letters, 38, p. 672

Somov B. V., Syrovatskii S. I. 1976, Soviet Physics Uspekhi, 19, p. 813

Struminskii A.B. 2011, Bulletin of the Russian Academy of Sciences: Physics, 
75, p. 751

The Catalogue of Solar Flare Events. 2008, //ww.wdcb.ru/stp/data/Solar

\_Flare\_Events/Fl\_XXIII.pdf.

Tylka A., Cohen W. F., Dietrich M. A. {\it{et al.}} 2005, ApJ, 625, p. 474
 
Zhang J., Dere K. P., Howard R. A., Kundu M. R., White S. M. 2001, Astrophys. J., 559, p. 452

\end{flushleft}

\end{document}